\documentclass[conference]{IEEEtran}
\IEEEoverridecommandlockouts
\usepackage{cite}
\usepackage{amsmath,amssymb,amsfonts}
\usepackage{algorithmic}
\usepackage{graphicx}
\usepackage{textcomp}
\usepackage{xcolor}

 \usepackage{mathrsfs}
\usepackage{algorithm}
\usepackage{booktabs}
\usepackage{float}
\usepackage{multirow}
\usepackage{amsthm}
\usepackage{enumitem}
\usepackage{subcaption}

\newtheorem{theorem}{Theorem} 

\newtheorem{problem}{Problem} 
\newtheorem{remark}{Remark}
\newtheorem{proposition}{Proposition}
\newtheorem{definition}{Definition}

\newtheorem{corollary}{Corollary}

\def\BibTeX{{\rm B\kern-.05em{\sc i\kern-.025em b}\kern-.08em
    T\kern-.1667em\lower.7ex\hbox{E}\kern-.125emX}}
\begin{document}

\title{Hybrid Design of Multiplicative Watermarking for Defense Against Malicious Parameter Identification\\
}

\author{Jiaxuan Zhang$^{*}$, Alexander J. Gallo$^{*}$ and Riccardo M. G. Ferrari$^{*}$
\thanks{$^{*}$Delft University of Technology, Delft, The Netherlands.
        {\tt\small \{j.zhang-42@student., a.j.gallo@, r.ferrari@\} tudelft.nl}}
\thanks{This paper has been partially supported by the AIMWIND project, which is funded by the Research Council of Norway under grant no. 312486.}
}

\maketitle

\begin{abstract}
Watermarking is a promising active diagnosis technique for detection of highly sophisticated attacks, but is vulnerable to malicious agents that use eavesdropped data to identify and then remove or replicate the watermark. In this work, we propose a hybrid multiplicative watermarking (HMWM) scheme, where the watermark parameters are periodically updated, following the dynamics of the unobservable states of specifically designed piecewise affine (PWA) hybrid systems. We provide a theoretical analysis of the effecs of this scheme on the closed-loop performance, and prove that stability properties are preserved. Additionally, we show that the proposed approach makes it difficult for an eavesdropper to reconstruct the watermarking parameters, both in terms of the associated computational complexity and from a systems theoretic perspective.
\end{abstract}


\section{Introduction}
\label{chap: 01: Introduction}

Modern Industrial Control Systems (ICS) often employ Information Technology (IT) hardware and software, in order to be more performant and reach greater flexibility and interoperability. This evolution has also exposed critical industrial infrastructures to cyber attacks \cite{tanBriefSurveyAttack2020}, which may lead to system level disruption and grave consequences for operators and the public. \emph{The design of secure CPSs is therefore imperative}, and the field of \textit{secure control} has emerged as an active area of research over the past decade \cite{sandberg2022secure}. 

A promising research direction corresponds to so-called \textit{active attack detection methods}  \cite{moPhysicalAuthenticationControl2015, miaoCodingSensorOutputs2014, ferrariSwitchingMultiplicativeWatermarking2021b, griffioenMovingTargetDefense2021, guoOutputCodingBasedDetectionScheme2021, ghaderiBlendedActiveDetection2021}, which do not rely solely on the knowledge of the plant dynamics, but actively modify the plant inputs or outputs to enhance attack detectability. For instance, in \cite{moPhysicalAuthenticationControl2015} an additive random watermark is added to the plant inputs, while the inclusion of \textit{measurement encoding matrices} is explored in \cite{miaoCodingSensorOutputs2014}. Furthermore, \cite{ferrariSwitchingMultiplicativeWatermarking2021b} proposed a multiplicative watermarking scheme for sensor outputs, while randomly generated parallel auxiliary systems are employed in \cite{griffioenMovingTargetDefense2021}. Often, these methods rely on matched mechanisms being present on both the plant and the controller side in order to generate, and then validate and possibly remove the signals added to the plant inputs or outputs.

Although active detection methods have been shown to improve detection capabilities against malicious agents injecting false data on the communication network, they do so under the assumption that attackers do not adapt their behaviour in response to the defence strategies. Indeed, if the additional security measures put in place for defence are successfully identified by the attacker, the injected data can be suitably adapted to evade detection. Different methods have been proposed as countermeasures to this, e.g., in \cite{miaoCodingSchemesSecuring2017,zhaiSwitchingWatermarkingbasedDetection2021,ferrariSwitchingMultiplicativeWatermarking2021b}, the parameters of the active diagnosis scheme are switched over time, thus changing the parameters that must be identified by an attacker to remain stealthy. On the other hand, methods like \cite{griffioenMovingTargetDefense2021} naturally resist identification, as parameters are randomly generated at each time step. Of these methods, \cite{miaoCodingSchemesSecuring2017,griffioenMovingTargetDefense2021} rely on pseudo-random number generators to produce new parameters, which must be synchronized at the plant and the controller sides. Furthermore, a switching mechanism is proposed in \cite{ferrariSwitchingMultiplicativeWatermarking2021b}, relying on an event-triggered strategy to define when to update the parameters of the multiplicative watermarking systems. In \cite{galloCryptographicSwitchingFunctions2022} a method based on the elliptic curve cryptography is proposed to further improve security for multiplicative watermarking (MWM).


The problem of maintaining control system states and parameters \emph{private} and protect from eavesdroppers received significant attention. Solutions based on Differential Privacy (DP)\cite{bottegalPreservingPrivacyFinite2017, katewaDifferentialPrivacyNetwork2020} rely on injecting additional noise into the system in order to mask sensitive information. In an ICS, anyway, the addition of privacy noise would act as a disturbance, thus affecting control performances. Another family of approaches involve the use of encrypted control \cite{stobbe2022fully}, which nevertheless can lead to possibly unacceptable time overheads, which could impact stability margins. 

In this paper we propose a novel active diagnosis method based on switching multiplicative watermarking \cite{ferrariSwitchingMultiplicativeWatermarking2021b}, where the switching law  is designed to resist identification. Specifically, we propose a \emph{hybrid multiplicative watermarking} (HMWM), where the watermark generator and remover switching law depends on the unobservable states of a suitably designed linear piecewise affine (PWA) dynamical system. This is instrumental in endowing our scheme with additional protections against eavesdropping attacks. 

The main contributions of this paper are as follows:
\begin{itemize}
    \item we propose a hybrid multiplicative watermarking method, providing a procedure to design a stable PWA hybrid generator and remover filters with unobservable states, and an example of switching function under which each mode is active with uniform probability;
    \item we demonstrate that, by exploiting the HMWM filters switching dynamics and switching rules, our method provides adequate guarantees that malicious agents cannot identify safety parameters.
\end{itemize}

The rest of this paper is organized as follows. In Section~\ref{chap: 02: problem formulation}, we formulate the problem, and define the system structure, the MWM filters, and the attacker capabilities. In Section~\ref{chap: 03: design of HMWM}, we propose our definition of the PWA HMWM system design for the filters, presenting the algorithm to be followed for parameter design. In Section~\ref{chap: 04: Identification Resistance}, we analyze the HMWM method's performance in resisting the identification from eavesdropping attackers. Finally, in Section~\ref{chap: 05: numerical example} we demonstrate our proposed HMWM scheme's effectiveness via numerical simulations.

\textit{Notation}:
Throughout the paper, the following notation is used. $\mathbb{Z}_+$ denotes the set of nonnegative integers. $I_n$ represents the $n$-dimensional identity matrix, while $0_{n\times m} \in \mathbb{R}^{n\times m}$ is a matrix of zeros; whenever clear from context, the subscripts $n$ and $n\times m$ are omitted.
Given a matrix $X \in \mathbb{R}^{n\times n}$, $\sigma(X)$ denotes its spectrum, and $\rho(X)$ its spectral radius.
A matrix $X \in \mathbb{R}^{n\times n}$ is said to be orthogonal, or orthonormal, if it is invertible and $X^{-1} = X^\top$. The space of symmetric matrices in $\mathbb R^{n\times n}$ is defined as $\mathbb S^n \subset \mathbb{R}^{n\times n}$.
For any two matrices $X_1$ and $X_2$, let $X = \operatorname{diag}(X_1, X_2)$ denote the block-diagonal matrix defined by $X_1$ and $X_2$.
The space of symmetric matrices in $\mathbb R^{n\times n}$ is defined as $\mathbb S^n \subset \mathbb{R}^{n\times n}$.
Notation $X \succ (\succeq) 0$ is used to state that a symmetric matrix $X \in \mathbb{S}^n$ is positive (semi)definite; similarly, a negative (semi)definite matrix is defined as $x \prec (\preceq) 0$.
Given a time-varying signal $x[k] \in \mathbb{R}^n$, $k \in \mathbb{Z}_+$, $x[k_1:k_2]$ is the sequence of instances $x[k], k \in \{k_1,k_1+1,\dots,k_2\}\subseteq \mathbb Z_+$.
A polyhedron $\mathscr X \subset \mathbb{R}^{n\times n}$ is a convex set, defined as $\mathscr{X} = \{x\in \mathbb{R}^n : Hx \leq k\}$, where $H\in \mathbb{R}^{m\times n}$ and $k \in \mathbb{R}^m$.
For any two sets $\mathscr{A}$ and $\mathscr{B}$, $\mathscr A \times \mathscr B$ denotes their Cartesian product.

\section{Problem Formulation}
\label{chap: 02: problem formulation}

\begin{figure}[ht]
    \centering
    \includegraphics[width = 0.9\linewidth]{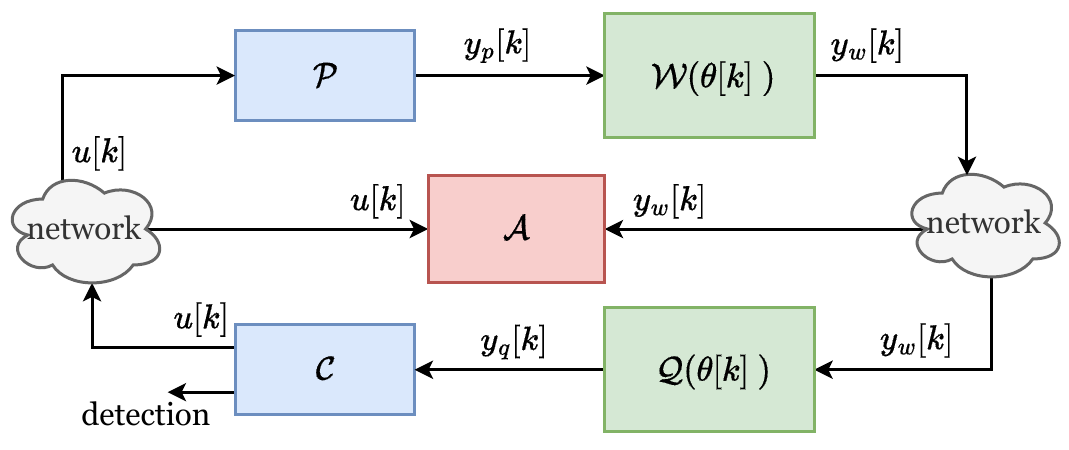}
    \caption{Overal System Description}
    \label{fig: 02 Overal System Description}
\end{figure}

We consider a cyber-physical system composed of a physical plant $\mathcal P$ and a controller $\mathcal{C}$, containing a steady-state Kalman filter, a static state feedback controller and an anomaly detector. The information between the controller and plant is exchanged over a communication network, thus exposing the CPS to attacks. To counteract this, we suppose the CPS is equipped with a switching multiplicative watermarking pair $(\mathcal W,\mathcal Q)$. The considered CPS structure is shown in Figure \ref{fig: 02 Overal System Description}.

\subsection{System Model}

The plant is modeled as an LTI system with dynamics
\begin{equation}\label{eq:sys}
\begin{aligned}
    x_p[k+1] &= A_p x_p[k] + B_p u[k] + w_p[k]; \\
    y_p[k] &= C_p x_p[k] + v_p[k]
\end{aligned}
\end{equation}
where $x_p \in \mathbb{R}^{n}, y_p\in\mathbb{R}^{p}$ are the plant's state and measurement output, and $u[k]\in\mathbb{R}^{m}$ is the control input. The signals $w_p \in \mathbb{R}^{n}$ and $v_p \in\mathbb{R}^{p}$ represent process and measurement noise, assumed to be realizations of identically and independently distributed zero-mean Gaussian processes with covariances $\Sigma_w \succ 0, \Sigma_v \succ 0$. We assume that all matrices are of appropriate dimensions, and that $(A_p, B_p)$ and $(C_p,A_p)$ are respectively controllable and observable pairs.

The controller ($\mathcal{C}$) is constituted of three components: a steady-state Kalman filter with observer gain $L$, a static state feedback controller with controller gain $K$ and a $\chi^2$ detector as an anomaly detector\footnote{Note that, although not the focus of this paper, we have included an anomaly detector, as multiplicative watermarking is predominantly a method for active attack diagnosis.}. These three components can be represented as the following dynamical system:
\begin{equation}\label{eq:ctrl}
    \begin{aligned}
        \mathcal{C}: \left\{
        \begin{aligned}
            \hat{x}_p[k+1] &= A_p \hat{x}_p[k] + B_p u[k] + L (y_p[k] - C_p \hat{x}_p[k])\\
            \quad u[k] &= - K  (\hat{x}_p[k] - x_{p, ref} )+ u_{ref}\\
            \quad r[k] &= y_p [k] -  C_p\hat{x}_p[k]
        \end{aligned}
      \right.   
    \end{aligned}
\end{equation}
where $\hat{x}_p \in \mathbb{R}^{n}$ is the estimated state, and $x_{p, ref} \in \mathbb{R}^{n}$, $u_{ref} \in \mathbb{R}^m$ are the reference state and control input, which are assumed to be piecewise constant.
Finally, $r \in \mathbb{R}^p$ is the Kalman filter innovation, which is used as a residual for attack diagnosis. The specific definition of the diagnosis tool is omitted from this paper, as out of its scope, but interested readers can turn to \cite{ferrariSwitchingMultiplicativeWatermarking2021b} for further details.

\subsection{Multiplicative Watermarking}
\label{chap: 02: MWM introduction}

Proposed in \cite{ferrariDetectionIsolationReplay2017b, ferrariSwitchingMultiplicativeWatermarking2021b}, switching multiplicative watermarking is an active technique for attack detection, whereby a watermarking generator ($\mathcal{W}$) filters $y_p$ before its transmission over the communication network to the controller. Once received, a suitably defined watermarking remover ($\mathcal{Q}$) then processes the information, returning a signal used by the controller. 

Let us start by defining the information available at the MWM generator and remover at time $k$ as follows:
\begin{equation}
    \begin{aligned}
    \mathscr{I}_w[k] &\triangleq\left\{y_w[0: k], y_p[0: k], x_w[0: k], \theta_w[0: k] \right\},\\
    \mathscr{I}_q[k] &\triangleq\left\{y_w[0: k], y_q[0: k], x_q[0: k], \theta_q[0: k],u[0: k]\right\}.
    \end{aligned}
\end{equation}

Both the watermarking generator and remover are time-varying systems, with dynamics described as follows:
\begin{equation}\label{eq:WM}
    \begin{aligned}
    & \mathcal{W}:\left\{\begin{aligned}
    x_w[k+1] & \!=\!A_w(\theta_w[k]) x_w[k]\!+\!B_w(\theta_w[k]) y_p[k] \\
    y_w[k] & \!=\!C_w(\theta_w[k]) x_w[k]\!+\!D_w(\theta_w[k]) y_p[k]
    \end{aligned} \right.\\
    & \mathcal{Q}:\left\{\begin{aligned}
    x_q[k+1] & \!=\!A_q\left(\theta_q[k]\right) x_q[k]\!+\!B_q\left(\theta_q[k]\right) y_w[k] \\
    y_q[k] & \!=\!C_q\left(\theta_q[k]\right) x_q[k]\!+\!D_q\left(\theta_q[k]\right) y_w[k] \\
    \end{aligned}\right. \\
    & \mathcal{F}: \theta_w[k] \!=\! f_w(\mathcal I_w[k]); \quad \theta_q[k] \!=\! f_q(\mathcal I_q[k])
    \end{aligned}
\end{equation}
where $x_w, x_q \in \mathbb{R}^{n_w}$ are the watermark generator and remover states, $y_w, y_q \in \mathbb{R}^p$ their outputs, and $\theta_w[k],\theta_q[k] \in \mathbb{R}^{n_\theta}$ are their parameters at time $k$, with $n_\theta = (n_w+p)^2$;
$f_w: \mathscr{I}_w \rightarrow \mathbb{R}^{n_\theta}$ and $f_q: \mathscr{I}_q \rightarrow \mathbb{R}^{n_\theta}$ are switching functions. We give the following definition of multiplicative watermarking pairs, following \cite{ferrariSwitchingMultiplicativeWatermarking2021b}.

\begin{definition}[Watermarking pair]\label{def:WM}
Two systems $(\mathcal W,\mathcal Q)$, with dynamics \eqref{eq:WM}, are said to be a watermarking pair if the following hold:
\begin{enumerate}[label=\alph*.]
    \item \label{def:WM:inv} $\mathcal{W}$ and $\mathcal{Q}$ are stable and invertible;
    \item \label{def:WM:eqParam} if $\theta_w[k] = \theta_q[k]$, $y_q[k] = y_p[k]$, i.e., ${\mathcal Q = \mathcal W^{-1}}$.
    $\hfill\triangleleft$
\end{enumerate}
\end{definition}

To meet Definition~\ref{def:WM}.\ref{def:WM:inv}, the system matrices for $\mathcal Q$ are defined as:
\begin{equation}\label{eq:WM:inv}
    \begin{aligned}
        D_q(\theta) &= D_w(\theta)^{-1}; A_q(\theta) = A_w(\theta) - B_w(\theta)D_w(\theta)^{-1} C_w(\theta); \\
        B_q(\theta) & = B_w(\theta) D_q(\theta); \quad C_q(\theta) = -D_q(\theta) C_w(\theta).
    \end{aligned}
\end{equation}
where $\theta = \theta_w[k] = \theta_q[k]$.

\subsection{Attacker Capabilities}
\label{chap: 02: problem formulation: atk}
In this paper, we suppose an eavesdropping attacker $\mathcal A$ monitors the information being transmitted over the communication network between the plant and the controller, as depicted in Figure~\ref{fig: 02 Overal System Description}. Specifically, we define the following threat model:
\begin{description}
    \item[System knowledge:] The attacker knows the parameters of the plant and controller models $\{A_p,B_p,C_p, L,K\}$.
    \item[Disclousure resources:] The attacker has direct access to signals $y_w$ and $u$ transmitted over the communication network. The set of information available to the attacker at time $k$ can be therefore defined as:
    \begin{equation}
        \mathscr{I}_a[k] \triangleq \{A_p,B_p,C_p,L,K,u[0:k],y_w[0:k]\}\,.
    \end{equation}
    Note that $\theta_w[0], \theta_q[0] \not\in \mathscr I_a$.
    \item[Attack objective:] The malicious agent attempts to reconstruct the multiplicative watermarking parameters $\theta_w[k]$ and $\theta_q[k]$ for all $k \geq K_{id}^a, k \in \mathbb{Z}_+$. Without loss of generality, $K_{id}^a = 0$.
\end{description}

\subsection{Problem Formulation}
\label{chap: 02: problem formulation:probFor}
The switching rules represented by $f_w$ and $f_q$ affect the difficulty for a malicious agent to identify the parameters of the multiplicative watermarking parameters. The switching rules we are to design should meet several requirements:
\begin{enumerate} [label=\textbf{R\arabic*}]
    \item \textbf{Fast Switching}: The mode should switch rapidly; \label{guideline 1}
    \item \textbf{Randomness}: The switching sequence should not be known in advance; \label{guidelines 2}
    \item \textbf{Synchronization}: $\mathcal{W}$ and $\mathcal{Q}$ should have synchronized modes, i.e., the mode should be chosen based on common information of $\mathscr I_w[k]$ and $\mathscr I_q[k]$. \label{guidelines 3}
\end{enumerate}

\begin{remark}
    Requirement~\ref{guideline 1} states that an objective of our solution is that it be fast switching. We set this requirement to avoid any design strategy that includes a minimum dwell time, as it has been shown to be beneficial for parameter identification, as is pointed out in Section~\ref{chap: 04: Identification Resistance}.
    $\hfill\triangleleft$
\end{remark}

Given the scenario presented in the previous subsections, we are now ready to formally introduce the problem we address in this paper:
\begin{problem}\label{prob:probFor}
Given a cyber-physical system \eqref{eq:sys}-\eqref{eq:ctrl}, equipped with a multiplicative watermarking scheme \eqref{eq:WM}, design the time-varying parameters $\theta_w, \theta_q$ such that:
\begin{enumerate}[label=\alph*.]
    \item \label{prob:probFor:WM} $(\mathcal W,\mathcal Q)$ is a watermarking pair, as per Definition~\ref{def:WM};
    \item \label{prob:probFor:stab} the CPS maintains closed-loop stability under switching;
    \item \label{prob:probFor:atk} an attacker with the information set $\mathscr I_a$ and capabilities defined in Section~\ref{chap: 02: problem formulation: atk} cannot exactly reconstruct $\theta_w[k], \theta_q[k]$, for all $k \geq K_{id}^a$, i.e. the time and data complexity to exactly identify the parameters can be arbitrarily large. In this paper, it relates to meeting the requirements \ref{guideline 1}-\ref{guidelines 3}.
    $\hfill\triangleleft$
\end{enumerate}
\end{problem}

\section{Design of Hybrid Multiplicative Watermarking}
\label{chap: 03: design of HMWM}

In this section, we give an overview of the design of the multiplicative watermarking scheme in~\eqref{eq:WM} as composed of hybrid systems with piecewise affine (PWA) dynamics. This proves to be beneficial when analyzing the method's resilience to attacks defined in Section~\ref{chap: 02: problem formulation: atk}, as shown in Section~\ref{chap: 04: Identification Resistance}.

\subsection{HMWM Structure}
\label{chap: 03 design of HMWM: HMWM Structure}
We propose a design strategy that defines the dynamics of $\mathcal W$ and $\mathcal Q$ as piecewise affine (PWA) linear switched systems. More precisely, the dynamics of $\mathcal W$ are\footnote{The dynamics of $\mathcal Q$ are analogous to \eqref{equa: PWA hybrid MWM}, substituting subscript $w$ with $q$, and changing the input from $y_p[k]$ to $y_w[k]$, and defining the system matrices following \eqref{eq:WM:inv}.
}:
\begin{equation}
\label{equa: PWA hybrid MWM}
    \begin{aligned}
    \mathcal{W}&:\left\{\begin{aligned}
    x_{w}[k+1] &  = \sum_{i=0}^N \beta_{w,i}\left(A_{w,i} x_{w}[k] + B_{w,i} y_{p}[k]\right) \\
    y_{w}[k] &  = \sum_{i=0}^N \beta_{w,i}\left(C_{w,i} x_{w}[k] + D_{w,i} y_{p}[k]\right) 
    \end{aligned} \right.\\
    \mathcal{F}&: 
    \theta_w[k] = \sum_{i=0}^N \beta_{w,i} \theta_i\,,\,\,
    \beta_{w, i} = \left\{\begin{aligned}
        &1, \,\, \text{if }  x_{w,u}[k] \in \mathscr P_i\\
        &0, \,\, \text{otherwise}
    \end{aligned}\right.
    \end{aligned}
\end{equation}
where subscript $i \in \mathcal N = \{1,\dots,N\}$ indicates one of $N$ \textit{modes} of operation, and the boolean variables
$\beta_{w,i}[k] \in \{0,1\}, \forall i \in \mathcal N$ are used to determine which mode is active at any given time. In the definition of the switching function, we rely on the evaluation of a logical rule, namely the evaluation of $x_{w,u} \in \mathbb P_i$. Here, $x_{w,u} \in \mathbb{R}^{n_u}$, defined in the following, is a portion of the state $x_w$ which is unobservable from $y_w$, and $\mathscr{P}_i \subset \mathbb{R}^{n_u}, i \in \mathcal{N}$ are a polyhedrons, which are defined such that $\bigcup_{i\in\mathcal{N}} \mathscr{P}_i = \mathbb{R}^{n_u}$. An example of such a partition can be seen in Figure~\ref{fig: switching region overview}. In order to guarantee that $\sum_{i\in\mathcal{N}} \beta_{w,i} = 1$, a necessary condition for \eqref{equa: PWA hybrid MWM} to be a PWA linear switching system, $\mathscr{P}_i \cap \mathscr{P}_j = \emptyset$ must hold for all $i, j \in \mathcal N, i \neq j$. Given that $\mathscr P_i$ are closed sets, this does not hold for adjacent partitions; therefore, an additional logical law must be applied. Specifically, we impose that, if $x_{w,u}[k] \in \mathscr{P}_i \cap \mathscr{P}_j, i,j \in \mathcal N, i\neq j$, $\beta_{w,i} = 1$ if and only if $i<j$. This guarantees that, \textit{in practice}, the partitioning of $\mathbb R^{n_u}$ is non-overlapping.

Matrices $A_{w,i},B_{w,i}, C_{w,i}, D_{w,i}$ are defined as follows:
\begin{equation}\label{eq:WM:Aw}
    \begin{matrix}
             A_{w,i} = \begin{bmatrix}
                 A_{w,i}^{-} &0 \\ 0 &A_{w,u}
             \end{bmatrix}\,,&  B_{w,i} = \begin{bmatrix}
                 B_{w,i}^{-}\\ B_{w,u}
             \end{bmatrix}\,, 
             \vspace{.1cm}\\
             C_{w,i} = \begin{bmatrix}
                 C_{w,i}^{-} &0
             \end{bmatrix}\,,
            & D_{w,i} = D_{w,i}^{-}\,.
    \end{matrix} 
\end{equation}
Note that, given \eqref{eq:WM:Aw}, the resulting systems are unobservable by definition, with  $x_{w,u} \in \mathbb{R}^{n_u}$ the unobservable portion of the state $x_w = \left[x_{w,o}^\top,x_{w,u}^\top\right]^\top$. In \eqref{eq:WM:Aw}, $A_w = \operatorname{diag}(a_{w,1},\dots,a_{w,n_u})$ and $B_{w,u} \in \mathbb{R}^{n_u\times p}$ are common to all modes, with $|a_{w,j}| \leq \sqrt{0.5}, \forall j \in \{1,\dots,n_u\}$; the matrices $A_{w,i}^-$ are defined as $A_{w,i}^- \triangleq \bar{T}^\top \bar{A}_{w,i}^- \bar{T}$, where $\bar{T}$ is an orthogonal matrix common to all modes, while $\bar{A}_{w,i}^-$ are randomly defined, stable diagonal matrix.  The matrices $B_{w,i}^-$ are defined randomly, such that $(A_{w,i}^-,B_{w,i}^-)$ is stable. Finally, $C_{w,i}^-$ and $D_{w,i}^-$ are defined as follows: firstly, a matrix $K_i$ stabilizing $A_{w,i}^- - B_{w,i}^-K_i$ is found satisfying 
\begin{equation}\label{eq:WM:Ki}
    \begin{aligned}
        \left[\begin{matrix}
            X & A_{w,i}^- X+B_{w,i}^- Z_i \\
            (A_{w,i}^- X+B_{w,i}^- Z_i)^{\top} & X
            \end{matrix}\right] & \succ 0 \\
                X \succ 0; \quad K_i = -Z_i X^{-1} \quad \forall i\in\mathcal{N},
    \end{aligned}
\end{equation}
then $D_{w,i}^-$ is defined randomly to be square and invertible matrix, and $C_{w,i}^- = D_{w,i}K_i$. The procedure for designing the watermarking matrices is summarized in Algorithm~\ref{alg: generate GUAS W and Q}. In the following we prove that this design procedure satisfies Problem~\ref{prob:probFor}.\ref{prob:probFor:WM} and Problem~\ref{prob:probFor}.\ref{prob:probFor:stab}.

\subsection{Generator and Remover Stability}

To prove closed-loop stability, we exploit the definition of globally uniformly asymptotic stability (GUAS) \cite{liberzonBasicProblemsStability1999}, using a uniform quadratic Lyapunov function \cite{fengStabilityAnalysisPiecewise2002}. 

\begin{algorithm} [t]
    \renewcommand{\algorithmicrequire}{\textbf{Input:}}
    \renewcommand{\algorithmicensure}{\textbf{Output:}}
    \caption{Generate GUAS $\mathcal{W}$ and $\mathcal{Q}$} 
    \label{alg: generate GUAS W and Q} 
    \begin{algorithmic}[1]
        \REQUIRE $n_w\geq 1,N\geq 1$
        \ENSURE $\theta_i, i\in\mathcal N$
        \STATE \label{stp:AT} Randomly generate diagonal matrices $\bar{A}_{w,i}^-$, ${i \in \mathcal N}$, such that  $\rho(\bar{A}_{w,i}^-) < 1$, and an orthonormal matrix $\bar{T}$ 
        \STATE \label{stp: Transform A}  Define $A_{w,i}^- = \bar{T}^\top \bar{A}_{w,i}^{-} \bar{T}$. 
        \STATE  \label{stp: B} Randomly generate $B_{w,i}^-$ such that $({A}_{w,i}^-, B_{w,i}^-)$ are controllable;
        \STATE \label{stp: K} Design $K_i$ such that \eqref{eq:WM:Ki} is jointly satisfied for all $i \in \mathcal N$;
        \STATE \label{stp:CD} Randomly generate $D_{w,i}$ and define $C_{w,i} = D_{w,i} K_i$.
        \STATE \label{stp:unobs} Randomly generate $a_w\in \mathbb{R} \quad \& \quad |a_w| \le \sqrt{0.5}, b_w^\top \in \mathbb{R}^{p}$, set $A_{w,u} = a_w$ and define $A_{w,i},B_{w,i},C_{w,i},D_{w,i}$ solving \eqref{eq:WM:Aw}.
        \STATE \label{stp: inv} Define $A_{q,i}$, $B_{q,i}$, $C_{q,i}$, $D_{q,i}$, corresponding to $A_{w,i}$, $B_{w,i}$, $C_{w,i}$, $D_{w,i}$,  solving~\eqref{eq:WM:inv};
        \STATE \label{stp:nW}
        \textbf{if} $n_w > 1$, \textbf{for} $t = 2:n_w$
        \STATE \quad Define: $$\begin{matrix}
            A_{w,i}^- = A_{w,i}\,, &B_{w,i}^- = B_{w,i}\,,\\
            C_{w,i}^- = C_{w,i}\,, &D_{w,i}^- = D_{w,i}\,;
        \end{matrix}$$
        \STATE \quad Repeat Step~\ref{stp:unobs}
        \STATE \textbf{endif} \textbf{endfor}
    \end{algorithmic} 
\end{algorithm}

\begin{theorem}\label{thm:stab}
    Given a watermark generator and remover pair $(\mathcal W,\mathcal Q)$ with dynamics as in \eqref{eq:WM:inv}, if their system matrices are generated following Algorithm~\ref{alg: generate GUAS W and Q}, with $n_u = 1$, the systems are GUAS and input-state stable (ISS) under arbitrary switching. Furthermore, $\mathcal{Q}(\theta_i) = \mathcal{W}(\theta_i)^{-1}, \, \forall i \in \mathcal N$.
    $\hfill\square$
\end{theorem}

\begin{proof}
    To prove that $\mathcal W$ is GUAS stable, it is sufficient to show that the autonomous systems
    \begin{equation}
        x_w[k+1] = A_{w,i}x_w[k]
    \end{equation}
    admit a common Lyapunov function for all $i \in \mathcal N$. We define a candidate Lyapunov function $V_w: \mathbb{R}^{n_w} \rightarrow \mathbb{R}$, $V_w(x) = x^\top P_w x$, where $P_w \succ 0, P_w \in \mathbb{S}^{n_w}$. Thus, it is sufficient that 
   \begin{equation}\label{eq:DeltaVw}
        A_{w,i}^\top P_w A_{w,i} - P_w \prec 0\,, \quad \forall i \in \mathcal N.
    \end{equation}

    Note that, given the definition of $A_{w,i}$ in \eqref{eq:WM:Aw}, a transformation $T = \operatorname{diag}(\bar{T},I_{n_u})$ can be defined such that $A_{w,i} = T^\top  \bar A_{w,i} T$, with $\bar{A}_{w,i} = \operatorname{diag}(\bar A_{w,i}^-,A_{w,u})$ and $T$ common to all modes. We therefore rewrite \eqref{eq:DeltaVw} as:
    \begin{equation}\label{eq:DeltaVw_1}
        T^\top \bar A_{w,i} T P_w T^\top \bar A_{w,i}^- T - P_w \prec 0\,, \quad \forall i \in \mathcal N.
    \end{equation}
    We now pre- and post-multiply \eqref{eq:DeltaVw_1} by $T$ and $T^\top$, respectively, and define $\bar P_w = T P_w T^\top \in\mathbb S^{n_w}$. Note that, because $P_w\succ 0$, $\bar P_w\succ 0$ as well. Thus, if
    \begin{equation}\label{eq:DeltaVw-}
        \begin{array}{cc}
            \bar A_{w,i}^{\top} \bar P_w \bar A_{w,i} - \bar P_w \prec 0
        \end{array}
    \end{equation}
    holds, so does \eqref{eq:DeltaVw}.
    Given $\bar A_{w,i}, \forall i \in \mathcal N$ by design is a diagonal matrix with $\rho(\bar A_{w,i}) < 1$, there exists a positive definite $\bar P_w$ such that \eqref{eq:DeltaVw-} holds for all $i\in \mathcal N$. This proves that $\mathcal W$ is GUAS under arbitrary switching, including the switching function in \eqref{equa: PWA hybrid MWM}.

    Similarly, we prove $\mathcal Q$ is GUAS, by supposing that there exists a symmetric $P_q \succ 0$ such that the candidate Lyapunov function $V_q: \mathbb R^{n_w} \rightarrow \mathbb R$, $V_q(x) = x^\top P_q x$ is suitable for all modes $i \in \mathcal N$. We prove this by construction. 
    Firstly, note that given definition of $K_i$ in \eqref{eq:WM:Ki}, each $A_{q,i}^- = A_{w,i}^- - B_{w,i}^- D_{w,i}^{-1}C_{w,i}^- = A_{w,i}^- - B_{w,i}^- K_i$ is Schur stable, and $V_q^- : \mathbb{R}^{n_w - n_u} \rightarrow \mathbb{R}$,    $V_q^-(x) = x^\top P_q^- x$ is a common Lyapunov function for all $i \in \mathcal N$, with $P_q^- = X^{-1}$, where $X\succ 0$ solves \eqref{eq:WM:Ki}. Furthermore, from definition of $A_{q,i}$ in \eqref{eq:WM:inv} and matrices in \eqref{eq:WM:Aw}, $A_{q,i}$ can be written as:
    \begin{equation}\label{eq:WM:AqDecomp}
        \begin{split}
            A_{q,i} &=\begin{bmatrix}
                A_{w.i}^- - B_{w,i}^- D_{w,i}^{-1}C_{w,i}^- &0\\
                -b_w D_{w,i}^{-1}C_{w,i}^- & a_w
            \end{bmatrix}\vspace{.1cm}\\ 
            &\triangleq \begin{bmatrix}
            A_{q,i,1} &0 \\
            A_{q,i,3} &A_{q,i,4}
        \end{bmatrix}
        \end{split}\,,
    \end{equation}
    with $a_w = A_{w,u} \in \mathbb{R}$, given $n_u = 1$ by assumption.
    Let us now introduce $p_q > 0, p_q \in \mathbb{R}$, and define $P_q = \operatorname{diag}(P_q^-,p_q)$. For $V_q$ to be an appropriate Lyapunov function, it is sufficient that
    \begin{equation}\label{eq:DeltaVq}
        A_{q,i}^\top P_q A_{q,i} - P_q \prec 0
    \end{equation}
    holds for all $i \in \mathcal N$. By considering the decomposition of $A_{q,i}$ defined in \eqref{eq:WM:AqDecomp}, we rewrite \eqref{eq:DeltaVq} as:
    \begin{equation}
        \begin{bmatrix}
            \Phi & \Xi\\
            \Xi^\top &\Psi
        \end{bmatrix}\prec 0,
    \end{equation}
    where
    \begin{equation*}
        \begin{split}
            &\Phi \triangleq A_{q,i,1}^\top P_q^- A_{q,i,1} - P_q^- + A_{q,i,3}^\top p_q A_{q,i,3}\\
            &\Psi \triangleq A_{q,i,4}^\top p_q A_{q,i,4} - p_q\\
            &\Xi \triangleq A_{q,i,3} p_q A_{q,i,4}
        \end{split}
    \end{equation*}
    Thus, by applying the Schur complement, \eqref{eq:DeltaVq} holds iff 
    \begin{subequations}
        \begin{gather}
            \Psi \prec 0; \label{eq:DeltaVq:condBr}\\
            \Phi - \Xi\Psi^{-1}\Xi^\top \prec 0.
            \label{eq:DeltaVq:condCompl}
        \end{gather}
    \end{subequations}
    By substituting the definition of $\Psi$ and $A_{q,i,4}$, \eqref{eq:DeltaVq:condBr} is equivalent to $(1-a_w^2)<0$, and therefore holds if and only if $|a_w| < 1$, which holds by construction. Furthermore, the design procedure in Algorithm~\ref{alg: generate GUAS W and Q} guarantees that \eqref{eq:DeltaVq:condCompl} holds: indeed, after some algebraic manipulations, recalling that $a_w, p_q \in \mathbb{R}$, we rewrite \eqref{eq:DeltaVq:condCompl} as:
    \begin{equation}
            P_q^- - A_{q,i}^{-\top} P_q^- A_{q,i}
            - A_{q,i,3}^\top A_{q,i,3} \frac{1-2a_w^2}{1-a_w^2} p_q \prec 0.
    \end{equation}
    Therefore, given that $(1-a_w^2) < 0$, it is necessary for $(1-2a_w^2)\geq0$, which holds if $|a_w| \leq \sqrt{0.5}$, corresponding to the constraint set in Step~\ref{stp:unobs} in Algorithm~\ref{alg: generate GUAS W and Q}. This completes the proof that $\mathcal Q$ is GUAS.

    Furthermore, $\mathcal W$ and $\mathcal Q$ are ISS, as $V_w$ and $V_q$ are contiunous uniform strict Lyapunov functions \cite{lazarSubtletiesRobustStability2007}.

    Finally, we complete the proof by pointing out that $\mathcal Q(\theta_i) = \mathcal W(\theta_i)^{-1}, \forall i \in \mathcal N$ holds by construction, via Step~\ref{stp: inv} in Algorithm~\ref{alg: generate GUAS W and Q}.
\end{proof}
\begin{corollary}\label{cor:LOG}
    The results of Theorem~\ref{thm:stab} hold for $n_u > 1$.
    $\hfill\square$
\end{corollary}
\begin{proof}
    Suppose that $n_u = 2$. Following the procedure indicated in Step~\ref{stp:nW} in Algorithm~\ref{alg: generate GUAS W and Q}, note that the value taken by $A_{w,i}^-$ in this proof is the same as that taken by $A_{w,i}$ in the proof of Theorem~\ref{thm:stab}. The same goes for all other matrices. Because $A_{w,i}$ is block diagonal, it is sufficient to define a new positive definite matrix $P_w = \operatorname{diag}(P_w^-,p_w) \succ 0$, where $P_w^-\succ 0$ is the matrix defining the Lyapunov function for $\mathcal W$ in the proof of Theorem~\ref{thm:stab}, and $p_w > 0, p_w \in \mathbb R$. Given that $|a_w| < \sqrt{0.5}<1$ by its definition in Step~\ref{stp:unobs}, 
    $$A_{w,i}^\top P_w A_{w,i} - P_w \prec 0$$
    holds for any $p_w > 0$. Thus $\mathcal W$ GUAS and ISS under arbitrary switching, for $n_u = 2$.
    On the other hand, following the same reasoning in the proof of Theorem~\ref{thm:stab}, given $|a_{w,2}| < \sqrt{0.5}$, GUAS and ISS of $\mathcal Q$ is proven.
    The proof for $n_u > 2$ follows by induction and recursive definition of $A_{w,u}$ in Algorithm~\ref{alg: generate GUAS W and Q}.
\end{proof}


\subsection{Switching Rule Design}
\label{chap: 03: design of HMWM: switching rule}

The switching rule in Section~\ref{chap: 03 design of HMWM: HMWM Structure} meets the requirements outlined in Section~\ref{chap: 02: problem formulation:probFor}. Indeed, 
    \ref{guideline 1} is met because, although exact quantification of the dwell time between switching events is challenging, the boundaries of each region $\mathscr P_i$ can be defined such that the probability of $x_{w,u}[k] \in \mathscr P_i$ is uniform across all $i \in \mathcal N$, given knowledge of the probability distributions of $w[k]$ and $v[k]$; 
    \ref{guidelines 2} is met, as the switching can be seen as being ``truly random'': the dynamics of $x_{w,u}$ depend on $w$ and $v$, which are the result of physical processes, and are not generated by a pseudo-random number generator\footnote{Note that at design stage random-number generators are necessary for the definition of the system parameters; this is done offline and does not clash with our statement here.}. Therefore, it is not possible to define the trajectory of $x_{w,u}[k]$ \textit{a priori}. 
    Finally, \ref{guidelines 3} holds, as State and parameter synchronization is proven in  Proposition \ref{prop: synchronization proposition}.

\begin{proposition}
\label{prop: synchronization proposition}
Suppose a CPS as in \eqref{eq:sys}-\eqref{eq:ctrl} is equipped with the HMWM scheme \eqref{equa: PWA hybrid MWM}. If $x_w[0] = x_q[0]$, and $\mathcal W$ and $\mathcal Q$ share the same $\mathscr{P}_i, \forall i \in \mathcal N$, then $\theta_w[k] = \theta_q[k], \forall k \geq 0$.
\hspace*{0pt}
$\hfill\square$
\end{proposition}

\begin{proof}
    Let us start this proof by supposing that, at some time $k = \kappa$, $x_w[\kappa] = x_q[\kappa]$; thus, by definition of the hybrid multiplicative watermarking scheme in \eqref{equa: PWA hybrid MWM}, $\theta_w[\kappa] = \theta_q[\kappa]$, as $x_w[\kappa] = x_q[\kappa] \in \mathscr{P}_{i}$. Dropping explicit dependence on the watermarking parameters, as they are matched, we write the one time-step difference equation of $x_{wq} = x_w - x_q$:
    \begin{equation}
        \begin{split}
            x_{wq}[\kappa + 1] 
            &\overset{\mathrm{(a)}}{=}(A_{w,i} - B_{q,i}C_{q,i})x_w[\kappa] - A_{q,i} x_q[\kappa] \\
            & \hspace{1cm}+(B_{w,i} - B_{q,i} D_{w,i}) y_p[\kappa]\\
            &\overset{\mathrm{(b)}}{=} A_{q,i} x_{wq}[\kappa]
        \end{split}
    \end{equation}
    where (a) holds by definition of the dynamics of $x_w$ and $x_q$, and     (b) holds by definition of the watermarking system matrices \eqref{eq:WM:inv}. Thus, $x_{wq}[\kappa+1] = x_{wq}[\kappa] = 0$, which in turn implies that $x_w[\kappa+1] = x_q[\kappa+1]$, and that $\theta_w[\kappa+1] = \theta_q[\kappa+1]$.
    The proposition's statement then holds by induction.
\end{proof}
\begin{remark}
    Let us note here that the switching law we present in this paper is different to the one presented in \cite{ferrariSwitchingMultiplicativeWatermarking2021b} in one fundamental aspect. Indeed, here the switching law at time $k$ depends on information available in $\mathscr I_w[k-1]$ and $\mathscr I_q[k-1]$. Instead, in \cite{ferrariSwitchingMultiplicativeWatermarking2021b}, the authors propose an event-triggered switching law, and the watermark remover must first \textit{decode} $y_w[k]$, then evaluate whether there has been a parameter jump in the watermark generator, and if that is the case, update its own parameters and recompute $y_q[k]$.
    $\hfill\triangleleft$
\end{remark}

\begin{proposition}
    The closed-loop of the CPS with watermarking pair $(\mathcal W, \mathcal Q)$ designed following Algorithm~\ref{alg: generate GUAS W and Q} is stable, and its performance remains unchanged, if $x_w[0] = x_q[0]$.
    $\hfill\square$
\end{proposition}
\begin{proof}
    The proof follows from the fact that, if designed following Algorithm~\ref{alg: generate GUAS W and Q}, $(\mathcal W, \mathcal Q)$ satisfy Definition~\ref{def:WM}, as shown in Theorem~\ref{thm:stab}, and their parameters match for all $k \in \mathbb Z_+$, as proven in Proposition~\ref{prop: synchronization proposition}.
\end{proof}

\subsection{Example Design of Switching Region}
\label{chap: 03 design of HMWM: Example Design of Switching Region}
Let us now propose a possible definition of the non-overlapping partitions $\mathscr{P}_i, i \in \mathcal N$. Specifically, we propose a partitioning of $\mathbb R^{n_u}$ such that, when the system reaches steady state, the probability of $x_{w,u}[k] \in \mathscr{P}_i$, at any $k$, is uniform across $i\in\mathcal N$. Let us start by characterizing the statistical properties of $x_{w,u}[k] \sim \mathcal N(\mu_{x_{w,u}}[k],\Sigma_{x_{w,u}}[k])$. From dynamics in \eqref{equa: PWA hybrid MWM}, we derive its mean and variance as:
\begin{equation}\label{eq:stats:unobs}
    \begin{split}
        &\mu_{x_{w,u}}[k] = A_{w,u} \mu_{x_{w,u}}[k-1] + B_{w,u} \mu_{y_{p}}[k-1]\\
        &\Sigma_{x_{w,u}}[k] = A_{w,u} \Sigma_{x_{w,u}}[k-1] A_{w,u}^\top + B_{w,u} \Sigma_{y_p}[k-1]B_{w,u}^\top
    \end{split}
\end{equation}
where $\mu_{y_p}[k]$ and $\Sigma_{y_p}[k]$ are the mean and variance of $y_p[k]$, in turn characterized by the mean and variance of $x_p[k]$:
\begin{equation}\label{eq:stats:sys}
    \begin{split}
        &\mu_{x_p}[k] = (A_p - B_pK) \mu_{x_p}[k-1] - B_pK \mu_e[k-1] \\
        &\mu_{y_p}[k] = C_p \mu_{x_p}[k] \\
        &\Sigma_{x_p}[k] = (A_p-B_pK) \Sigma_{x_p}[k-1] (A_p-B_pK)^\top \\
        & \hspace{2.5cm} + B_pK \Sigma_e[k-1] B_pK^\top + \Sigma_w\\
        &\Sigma_{y_p}[k] = C_p \Sigma_{x_p}[k] C_p^\top + \Sigma_v
    \end{split}
\end{equation}
where $\mu_e[k]$ is the expectd value of the estimation error $e[k] = x_p[k] - \hat{x}_p[k]$, and $\Sigma_e[k]$ its variance. By definition of the Kalman filter in \eqref{eq:ctrl}, $\mu_e[k] = 0$, while the steady-state definition of  $\Sigma_e[k]$ can be found in \cite{murguia2016cusum}. Given that $A_p - B_pK$ is Schur stable by design of $K$, it is possible to use \eqref{eq:stats:unobs}-\eqref{eq:stats:sys} to define the steady state values of $\mu_{x_{w,u}}$ and $\Sigma_{x_{w,u}}$. The steady-state statistics of $x_{w,u}$ can then be used to partition $\mathbb R^{n_u}$ into $N$ polyhedra, each having the same probability, using, e.g., the cumulative distribution function of the multiparametric Gaussian distribution.

\begin{remark}
    The procedure outlined in this section only considers using $x_{w,u}[k]$ as the decision variable of mode selection. This is, of course, only one possible solution, as mode selection can also depend on $x_w[k]$ as a whole, or $u[k]$. The evaluation of whether there are any (dis)advantages in making one choice instead of another is left for further work.
    $\hfill\triangleleft$
\end{remark}
\begin{remark}
    Note that the procedure proposed in this section to define $\mathscr{P}_i, i \in \mathcal N$ depend on the references $x_{p,ref},u_{ref}$, as it biases the unobservable state's mean. As such, it is necessary to change $\mathscr{P}_i$ whenever $x_{p,ref}$ changes, which requires $x_{p,ref}$ to be transmitted between $\mathcal C$ and $\mathcal P$, and for $\mathcal W$ to have sufficient computational resources to execute the computation. We leave the development of a definition of the partitioning $\mathscr{P}_i$ that is time-invariant as future work.
    $\hfill\triangleleft$ 
\end{remark}


\section{Identification Resistance}
\label{chap: 04: Identification Resistance}

Having presented our proposed design strategy for the HMWM in Section~\ref{chap: 03: design of HMWM}, and having thus addressed Problem~\ref{prob:probFor}.\ref{prob:probFor:WM} and Problem~\ref{prob:probFor}.\ref{prob:probFor:stab}, we can now evaluate our scheme against an adversarial eavesdropper attack, as the one defined in Section~\ref{chap: 02: problem formulation: atk}. Before providing details on this, note that, if seen from the perspective of cryptography, $\mathcal W$ and $\mathcal Q$ can be seen as procedures that encode and decode the transmitted data. From this viewpoint, $\theta_w[k]$ and $\theta_q[k]$ can be seen as secret keys, guaranteeing security. In assessing the security of cryptographic algorithms, the computational complexity required to \textit{break} them is evaluated, which often takes the form of evaluating the complexity of solving inverse problems over the field of integers modulo a prime \cite{katz2020introduction}. 
The techniques for evaluating the security of cryptographic algorithms inspire our evaluation of our proposed methodology, which relies on three metrics: 
\begin{enumerate}[label=\roman*.]
    \item the computational complexity of identifying the system parameters;
    \item the amount of memory required to perform identification;
    \item an evaluation of the theoretical difficulties associated with identifying the model of PWA hybrid system dynamics with unobservable states.
\end{enumerate}
In the remainder of this section, we demonstrate how, by designing the dynamics of $\mathcal W$ and $\mathcal Q$ according to Algorithm~\ref{alg: generate GUAS W and Q}, the obtained result is hard to identify.


\begin{theorem}\label{thm:cplx}
    Considering multiplicative watermarking systems $\mathcal W$ and $\mathcal Q$, designed following Algorithm~\ref{alg: generate GUAS W and Q}, the computational complexity of exactly identifying $\theta_w[k]$ and $\theta_q[k]$ from $\mathscr I_a[k]$ is $\mathcal{NP}$-hard.
    $\hfill\square$
\end{theorem}
\begin{proof}
    The proof follows directly from the computational complexity of solving exact identification of piecewise affine regression problems \cite[Ch.5]{lauer2019hybrid}
\end{proof}

\begin{remark}
    In \cite[Ch.5]{lauer2019hybrid}, analysis of the complexity of different bounded-error identification strategies for switched systems is conducted by restricting solutions to the set of rational numbers, rather than the reals. We apply these results here without loss of generality, as in practice the solution we propose is to be applied to a digital control system, and for matching parameters to be guaranteed, a fixed point representation is likely to be necessary.
    $\hfill\triangleleft$
\end{remark}

Although Theorem~\ref{thm:cplx} gives a result for the computational complexity of \textit{exact} identification of the system parameters, there are some methods to find some approximate solutions for input-output models of the system, such as the piecewise auto-regressive model with extra input (PWARX). One possible method is piecewise affine regression \cite{lauer2019hybrid}. The following result pertains to the difficulty of identifying PWA systems with unobservable outputs.

\begin{theorem}\label{thm:infDim}
    Consider a multiplicative watermarking scheme for which $\mathcal W$ and $\mathcal Q$ are designed following Algorithm~\ref{alg: generate GUAS W and Q}, does not admit a PWARX model.
    $\hfill\square$
\end{theorem}
\begin{proof}
    Given $\mathcal W$ and $\mathcal Q$ are defined to be unobservable, they can only be observed in infinite time \cite[Prop.3.1]{paolettiInputOutputRepresentationPiecewise2010}. The theorem's statement follows directly.
\end{proof}

\begin{remark}
    Note that there are some methods presented in literature to identify state-space models directly, e.g., \cite{lauer2019hybrid, bakoIdentificationSwitchedLinear2009, sefidmazgiSwitchedLinearSystem2015, lopesNewAlgorithmIdentification2013}. However, \cite{sefidmazgiSwitchedLinearSystem2015, lopesNewAlgorithmIdentification2013} assume a minimum dwell time, not satisfied by the scheme presented in this paper. Additionally, while the method in \cite{bakoIdentificationSwitchedLinear2009} does not require minimum dwell time, it does require the system to be pathwise-observable, again not a feature of the result of Algorithm~\ref{alg: generate GUAS W and Q}.
    $\hfill\triangleleft$
\end{remark}

Finally, let us comment on the storage space which may be necessary to compute an approximate solution to the parameters of the HMWM scheme. For this, we give a lower bound value, based on the minimum number of data points required to ensure the system is persistently excited (PE).

\begin{theorem}\cite[Thm. 2]{muPersistenceExcitationIdentifying2022}
    To ensure the regressors and corresponding membership indices are PE for the [switched linear] system, the minimum number of required samples is
    \begin{equation}
        \begin{aligned}
            \frac{(n_\theta-1) N^2+(n_\theta+1) N}{2}.
        \end{aligned}
    \end{equation}
    $\hfill\square$
\end{theorem}

Furthermore, in Table~\ref{tab: IO Identificaiotn Complexity} we include a characterization of the sample complexity required to perform identification using an approximate IO model, which truncates the input-output data at a horizon length of $n_h$, while identifying a state-space model with $n_m$ modes. In Table~\ref{tab: numerical: IO complexity} we show how this grows to be intractable as the horizon length increases.

Instead of identifying an infinite-dimension IO model, a learning attacker can use a finite-dimension IO model to approximate the switching dynamics. According to \cite{paolettiInputOutputRepresentationPiecewise2010, muPersistenceExcitationIdentifying2022} for a state-space model with $s$ submodels ($n_m = s$) and use $\nu$ horizons ($n_h = \nu$) to approximate it, the dimension of the IO model and sample complexity is shown in the Table \ref{tab: IO Identificaiotn Complexity}. The complexity grows intractable as the horizon and the number of modes grows.

\begin{table}[t]
\centering
\caption{IO Identification Complexity}
\label{tab: IO Identificaiotn Complexity}
\begin{tabular}{@{}ccccc@{}}
\toprule
$n_m$    & $n_h$   & IO        & IO dimension      & Sample Complexity \\ \midrule
 $s$   & $\nu$       & $s^{\nu}$     &   $(p\!+\!m) \nu$   & $\frac{((p \!+\! m)\nu \!-\!1)s^{\nu} \!+\! ((p \!+\! m)\nu\!+\!1)s^{\nu} }{2}$              \\ \bottomrule
\end{tabular}
\end{table}

\section{Numerical Example}
\label{chap: 05: numerical example}

\subsection{Overview}
\label{chap: 05 Numerical Example Overview}

We use the linearized quadruple-tank water system in \cite{ghaderiBlendedActiveDetection2021} as our test bench. The noise parameter, the linearized operating points, and the controller parameters we used  are as follows:
\begin{equation}
    \begin{aligned}
        x_{\mathrm{ref}}& =[5,5,2.044,1.399]^{\top}, \quad
        u_{\mathrm{ref}}=[0.724,1.165]^{\top} \\
        \mu_w &= \left[\begin{matrix}
         0 & 0 & 0 & 0
        \end{matrix}\right]^{\top}, \quad
        \mu_v = \left[\begin{matrix}
         0 & 0
        \end{matrix}\right]^{\top}, \\
        \Sigma_w &= 10^{-3} I_4, \quad
        \Sigma_v = 10^{-1} I_2 \\
        K & = \left[\begin{matrix}
            -3.0993 & -4.0721 & 2.0528 & -2.8417 \\
            -3.9353 & -3.3330 & -2.8461 & 1.9997
        \end{matrix}\right]\\
    \end{aligned}
\end{equation}

\subsection{HMWM Design and Simulation Overview}
\label{chap: 05: HMWM Design and Simulation Overview}
We designed $\mathcal{W}$ and $\mathcal{Q}$ with 5 states $(n_w = 5)$, of which 2 are unobservable  states $(n_{u} = 2)$. The number of modes is 6 ($N = 6$). We randomly generate 50 sets of HMWM parameters, and the simulation duration is 1000 steps for each set.
One set of parameters of the hybrid multiplicative watermarking generator's unobservable state is as follows:
\begin{equation}
    \begin{aligned}
        A_{w, u} & = \left[\begin{matrix}
            0.3908 & 0 \\ 
            0 & 0.6076
        \end{matrix}\right], 
        B_{w, u} = \left[\begin{matrix}
            0.1299 & 0.4694 \\ 
            0.5688 & 0.0119
        \end{matrix}\right],  \\
    \end{aligned}
\end{equation}

Following the example design procedure in section \ref{chap: 03 design of HMWM: Example Design of Switching Region}, the mean and variance of the steady-state $x_{w,u}$ are as follows. 
\begin{equation}
    \begin{aligned}
    \mu_{x_{w,u}} = [0.9838, 1.4800]^\top,\,
        \Sigma_{x_{w,u}} &= 
        \left[\begin{matrix}
            0.0283 & 0.0105 \\
            0.0105 & 0.0519
        \end{matrix}\right].
    \end{aligned}
\end{equation}


\begin{figure}
    \centering
    \begin{subfigure}{0.48\linewidth}
    \centering
    \includegraphics[width=\linewidth]{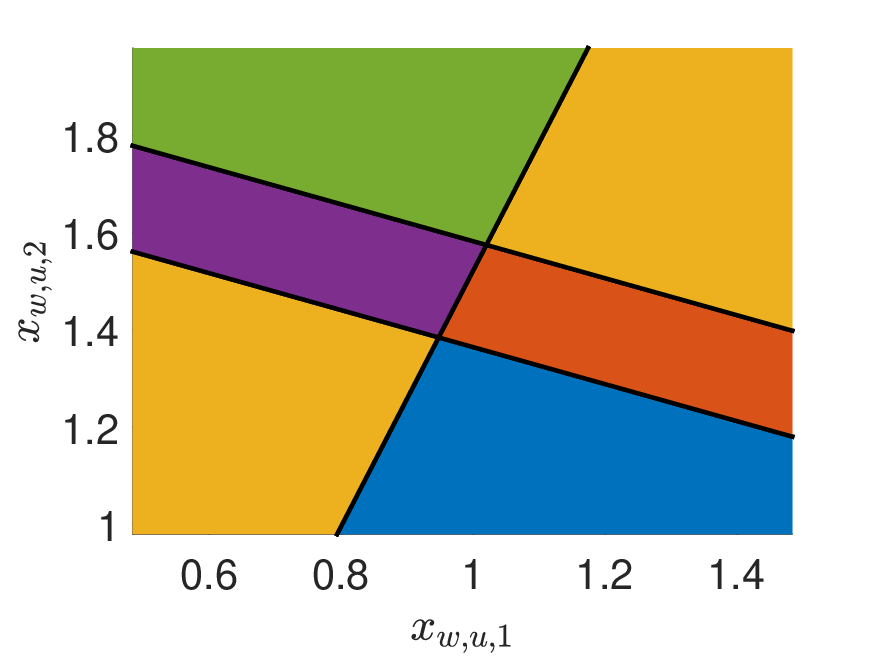}
    \caption{}
    \label{fig: switching region overview}
    \end{subfigure}
    \begin{subfigure}{0.48\linewidth}
      \centering
      \includegraphics[width=\linewidth]{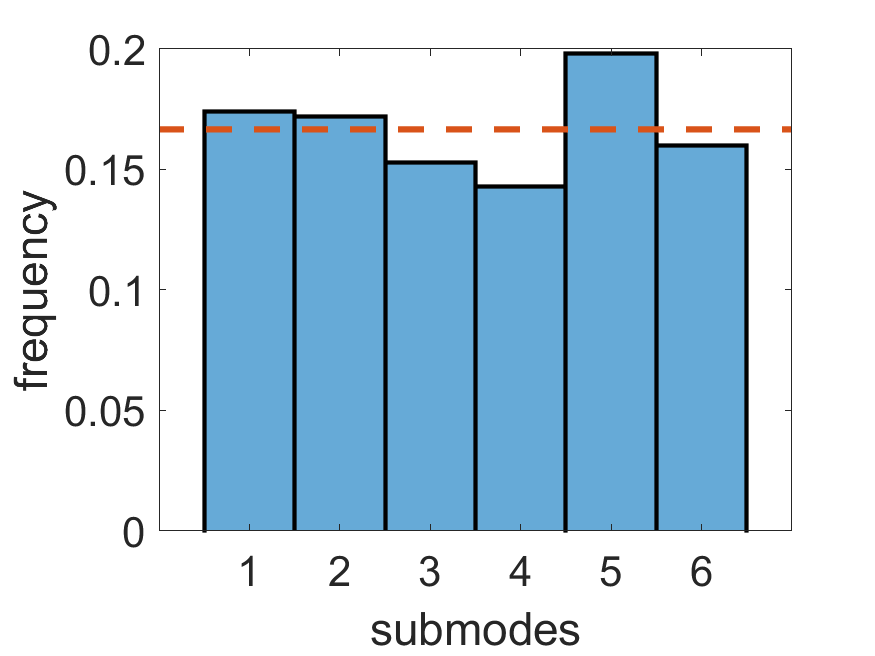}
      \caption{}
      \label{fig: rule fast switching}
    \end{subfigure}
    \caption{(a) partitioning of $\mathbb{R}^{n_u}, n_u = 2$ into $N = 6$ polyhedrons, each associated to a parameter $\theta_i, i \in \mathcal N$; 
    (b) relative frequency of each of the $N = 6$ modes over 1000 time steps.
    }
\end{figure}


For the given watermarking parameter, the IO model dimension and the minimum number of samples needed to meet the PE requirement for different horizon numbers are shown in Table \ref{tab: numerical: IO complexity}. The number of IO models and samples needed becomes intractable as the horizon number grows. 

\begin{table}[b]
\centering
\caption{Numerical IO Identification Complexity}
\label{tab: numerical: IO complexity}
\begin{tabular}{@{}ccccc@{}}
\toprule
 Horizon Number   & IO modes                  & number of samples  \\ \midrule
  1         & 6                         &     69        \\ 
  5         & 7776                      &  $5.7451 \times 10^8$\\
  10        &  $6.0466 \times 10^7$     &  $7.1295  \times 10^{16}$    \\
  15        &  $4.7018 \times 10^{11}$    &  $6.5217 \times 10^{24}$ \\ \bottomrule
\end{tabular}
\end{table}

\subsection{Performance}
\begin{figure}[t]
    \centering
    \includegraphics[width=1.0\linewidth]{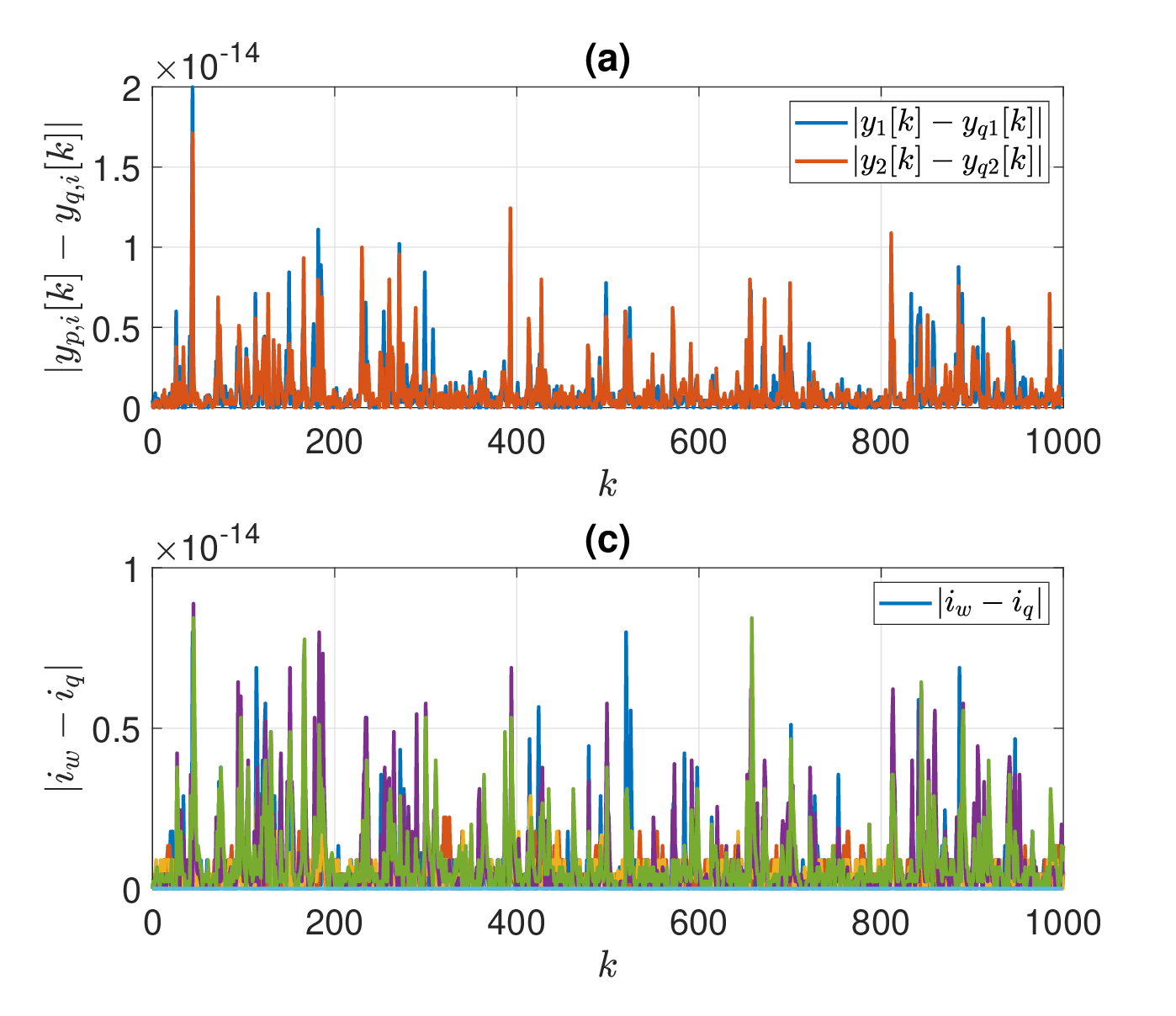}
    \caption{Switching Synchronization: (a) the absolute difference between the plant output and $\mathcal{Q}$ output; (b) the absolute difference between $\mathcal{W}$ states and $\mathcal{Q}$ states. 
    } 
    \label{fig: switching synchronization}
\end{figure}

The simulation result shows that the switching rule meets the requirements in section \ref{chap: 03 design of HMWM: HMWM Structure}: 
    \ref{guideline 1}: in Figure \ref{fig: rule fast switching} we show that, for $N = 6$ modes, each partition (shown in Fig.~\ref{fig: switching region overview}) each mode is active approximately the same amount of time; furthermore during this simulation, 637 switching events occur, the median dwell time is 1, with a maximum dwell time of 8. Performing the simulation with $N = 50$ modes, there are 588 switching events, with a median dwell time of 1.
    \ref{guidelines 2}: The randomness of the mode sequence is guaranteed by design.
    \ref{guidelines 3}: in Figure~\ref{fig: switching synchronization} we show synchronization error of the watermarking systems' states, outputs and parameters; although there is a small error in the states of $\mathcal W$ and $\mathcal Q$, as well as between $y_p$ and $y_q$ (cfr. Figure~\ref{fig: switching synchronization}.a-Figure~\ref{fig: switching synchronization}.b) this does not impact the mode selection. These errors can be ascribed to numerical errors in MATLAB.

Since all regression methods should finally match each data point to specific modes, table \ref{tab:  numerical: classification performance} shows the result of comparing the actual modes and labels estimated by different methods: k-means,  k-LinReg \cite{Lauer13klinreg}. The label is estimated based on IO data, and different horizons' results are presented. We choose the random index (RI), the Fowlkes–Mallows index (FMI), and the Jaccard index (JI) to measure the performance. The closer these indices to 1, the better the clustering result.

\begin{table}[t]
\centering
\caption{numerical: classification performance}
\label{tab:  numerical: classification performance}
\begin{tabular}{ccccc}
\hline
methods                   & Horizon & RI         &     JC           & FMI \\ \hline
\multirow{3}{*}{k-means}  & 1       & 0.3136     &     0.0476        & 0.2054    \\
                          & 2       & 0.9513     &     0.2433       &  0.3916   \\
                          & 3       & 0.2017     &      0.0386      & 0.1792    \\ \hline
\multirow{2}{*}{k-linreg} & 1       &  0.70867     &    0.1123       &   0.2027  \\
                          & 2       &   0.9264    &     0.0342      &  0.0662   \\
                          \hline
\end{tabular}
\end{table}

\section{Conclusion}
\label{chap: 06: conclusion}
In this work, we propose a piecewise-affine hybrid design for multiplicative watermarking. We provide methods to design parameters which guarantee the multiplicative watermarking systems are stable and invertible, and present a way of partitioning the state space such that the resulting switching is fast and randomic, while maintaining synchronization. We demonstrate the hardness that our proposed methodology offers against eavesdropping attacks.

In the future, we will focus on the detection performance of the proposed method, as well as comparing a number of different design choices when evaluating sensitivity to data injection attacks. Finally, we propose to extend our design algorithm to provide robustness against parameter mismatching.



\bibliographystyle{ieeetr}
\bibliography{ref}

\end{document}